\documentclass[review,number,sort&compress]{elsarticle}

\usepackage{lineno,hyperref}
\modulolinenumbers[1]
\usepackage{amsmath}
\usepackage{amsfonts}
\usepackage{amssymb}
\usepackage{rotating}
\usepackage{multirow}

\journal{Nucl. Instr. Meth. Phys. Res. A}

\begin{document}

\begin{frontmatter}

\title{On the performance of Zero Degree Calorimeters in detecting multinucleon events}

\author[MIPT,INR]{Uliana Dmitrieva}
\author[INR]{Igor Pshenichnov\corref{correspondingauthor}}

\cortext[correspondingauthor]{Corresponding author}
\ead{pshenich@inr.ru}

\address[MIPT]{Moscow Institute of Physics and Technology, Institutskiy Pereulok 9, Dolgoprudny, Moscow Region 141701, Russia}
\address[INR]{Institute for Nuclear Research of the Russian Academy of Sciences, Prospekt 60-letiya Oktyabrya 7a, Moscow 117312, Russia}

\begin{abstract}
The facilities designed to study collisions of relativistic nuclei, such as the MPD at NICA (JINR), STAR at RHIC (BNL), ALICE, ATLAS and CMS at the LHC (CERN), are equipped with pairs of hadronic Zero Degree Calorimeters (ZDC) to detect forward nucleons at the both sides of the interaction point and estimate the collision centrality. The energy deposited in a ZDC fluctuates from one event to another, but on average it is proportional to the number of absorbed nucleons. Forward nucleons are also emitted in electromagnetic dissociation (EMD) of nuclei in ultraperipheral collisions, and they are used to monitor the luminosity.  As known, ZDC energy spectra are specific to each facility, because they are affected by the ZDC acceptance, and the ZDC energy resolution depends on the beam energy. In this work a simple probabilistic model leading to handy formulas has been proposed to connect the numbers of emitted and detected forward nucleons taking into account a limited ZDC acceptance. The ZDC energy spectra from the EMD with the emission of one, two, three and four forward neutrons and protons have been modeled for the collision energies of NICA and the LHC. The case of a rather small ZDC acceptance has been investigated and a possibility to measure the inclusive nucleon emission cross section has been demonstrated. 
\end{abstract}


\end{frontmatter}


\section{Introduction}

There exists a relationship between the impact parameter as an important initial condition of a nucleus-nucleus collision event and the number of spectator nucleons beyond the overlap zone which continue to propagate in the forward direction after the collision. This motivates the use of forward hadronic calorimeters in studies of interactions of relativistic nuclei. In particular, the experiments at heavy-ion colliders like the MPD at NICA (JINR)~\cite{Golubeva2013,Golubeva2017}, STAR at RHIC (BNL)~\cite{Adler2001},  ALICE~\cite{Arnaldi2006,Puddu2007,Gemme2009}, ATLAS~\cite{White2010} and CMS~\cite{Grachov2011} at the LHC (CERN) are equipped with pairs of hadronic Zero Degree Calorimeters (ZDC) for detecting forward nucleons at the both sides of the interaction point. One of the most reliable methods to sort collision events into centrality classes is based on detecting spectator neutrons in the ALICE ZDC~\cite{Abelev2013o}. The electromagnetic dissociation (EMD) of nuclei in ultraperipheral collisions is another source of forward nucleons. This process is used to monitor the collider luminosity~\cite{Baltz1996,Pshenichnov2011} on the basis of the EMD cross sections which were reliably calculated~\cite{Pshenichnov2011,Pshenichnov2001} and accurately measured~\cite{Abelev2012n}.    

It is quite common to calculate the distributions of energy absorbed in a ZDC (ZDC energy spectra) by Monte Carlo modeling specifically for each facility. The simulations account for the actual beam energy, the geometric acceptance of ZDC and the efficiency of nucleon registration, which vary from one set-up to another. Therefore, ZDC energy spectra calculated for different facilities differ from each other, but nevertheless one can point out common characteristics of the spectra and study their dependence on the beam energy and ZDC acceptance. ZDC are designed for counting forward nucleons resulting from nucleus-nucleus collisions on the basis of energy deposited by these nucleons in ZDC. However, the performance of ZDC deteriorates when some of nucleons emitted in a multinucleon event do not hit ZDC.  Since spectator nucleons are emitted close to the directions of the colliding beams in heavy ion colliders~\cite{Adler2001,Arnaldi2006,Gemme2006}, the space available for placing a ZDC is rather limited.  A ZDC installed in a close proximity of beam pipes can be also partially obscured by collimators, vacuum chambers or other collider components~\cite{Gemme2006}. Forward protons hit the LHC beam pipes before they reach the ALICE  proton ZDC~\cite{Arnaldi2006,Gemme2006} and some protons are scattered by the walls of the pipes at large angles. In the present work a simple probabilistic model leading to handy formulas is proposed to relate the numbers of emitted and detected forward nucleons taking into account a limited ZDC acceptance.

\section{ZDC response to forward nucleons}\label{ZDC_response}

A ZDC is typically build in such a way that its dimensions are sufficient for the absorption of a primary forward nucleon as well as most of secondary particles created by this nucleon in electromagnetic processes and nuclear reactions inside the calorimeter~\cite{Puddu2007}. Therefore, the average energy deposited in the ZDC by a single spectator nucleon corresponds to its energy which, in its turn, amounts to the beam energy. The energy deposited in the ZDC fluctuates from one multinucleon event to another, but on average it is proportional to the number of absorbed nucleons. It is quite common to characterize the distribution of energy in the ZDC for one-nucleon events by means of a Gaussian with the mean $\mu_1$ equal to the beam energy $E_0$ and the dispersion $\sigma_1$ also depending on $E_0$. Two functions are usually considered to approximate the dependence of energy resolution $\sigma_1/\mu_1$ on $E_0$. For example, in Refs.~\cite{Golubeva2013,Arnaldi2006} the energy resolution has been evaluated as:
\begin{linenomath*}
\begin{equation}
\cfrac{\sigma_1}{\mu_1}=\sqrt{\cfrac{a^2}{E_0}+b^2} \ , 
\label{wsqrt}
\end{equation}
\end{linenomath*}
\noindent while in Refs.~\cite{Puddu2007,Awes1989} a bit different approximation has been adopted:
\begin{linenomath*}
\begin{equation}
\cfrac{\sigma_1}{\mu_1}=\cfrac{c}{\sqrt{E_0}}+d \ . 
\label{wosqrt}
\end{equation}
\end{linenomath*}
Naturally, the functions (\ref{wsqrt}) and (\ref{wosqrt}) are nearly equivalent to each other in the case of $a\approx c$ and the smallness of the second terms in comparison to the first ones at low beam energy.  The higher the beam energy, the better the ZDC energy resolution is.  For example, $\sigma_1/\mu_1$ calculated at $E_0=2510$~GeV with Eq.~(\ref{wsqrt}) for the ALICE neutron ZDC with the parameters $a=256.6$\%${\rm GeV}^{1/2}$ and $b=10.3$\% amounts to 11.5\% ~\cite{Arnaldi2006}. 

The numbers of forward nucleons are obtained in ALICE~\cite{Puddu2007,Abelev2012n} and other experiments~\cite{White2010,Grachov2011} by fitting the measured distributions of energy $E$ deposited in calorimeters in multinucleon events. In particular, the fitting functions $F(E)$ are constructed as the sum of four Gaussians corresponding to $i=1,2,...4$ nucleons emitted in EMD events~\cite{Abelev2012n}: 
\begin{linenomath*}
\begin{equation}
F(E)=\sum_{i=1}^{4} f_i(E)=\sum_{i=1}^{4} \frac{{\mathsf N_i}}{\sqrt{2\pi}\sigma_i} e^{-\frac{(E-\mu_i)^2}{2\sigma_i^2}} \ .
\label{Eq:FitFunc}
\end{equation}
\end{linenomath*}
Each Gaussian $f_i(E)$ representing an $i$-th peak is characterized by its mean value $\mu_i$, its dispersion $\sigma_i$ and the normalization constant $\mathsf N_i$ which is proportional to the numbers of events with $i$ nucleons. Here $\mu_1=E_0$, $\mu_i = i \mu_1$ and $\sigma_i = \sqrt{i} \sigma_1$.   In addition, a correction for the pedestal in the ZDC signal has been  introduced in Ref.~\cite{Abelev2012n}, which affects $\sigma_i$.  However, for the sake of simplicity the function~(\ref{Eq:FitFunc}) without a pedestal correction is used in the present work to represent the ZDC energy spectra in NICA/MPD and ALICE experiments.

\section{Correction for ZDC acceptance to the measured yields }\label{acc_corr}

As discussed above in Sec.~\ref{ZDC_response}, the numbers of events with different multiplicities of forward nucleons $\mathsf N_i$ can be reliably measured by fitting the ZDC energy distribution by the sum of Gaussians providing that all such nucleons are intercepted by the ZDC. However, the determination of $\mathsf N_i$ is not straightforward in the case when some of forward nucleons are lost due to a limited ZDC acceptance. These nucleons either do not hit the calorimeter at all or deposit a reduced energy due to their peripheral impact on the ZDC and shower leakage. In particular, in some of three-nucleon events either one or two nucleons can be lost.  As a result, such three-nucleon events are misidentified, respectively, as two-nucleon or one-nucleon events. In general, $\mathsf n_i$ as numbers of {\em detected} events of each nucleon multiplicity $i$ have to be used in Eq.~(\ref{Eq:FitFunc}) instead of {\em true} numbers $\mathsf N_i$.

The corrections for the ZDC acceptance and efficiency to the yields of one-, two- and three-neutron events measured in the EMD of 158A~GeV indium nuclei in collisions with Al, Cu, Sn and Pb targets has been introduced in Ref.~\cite{Karpechev2014}. Such corrections were specific to the experiment of Ref.~\cite{Karpechev2014}, but one can think of a more general approach to account for the ZDC acceptance.  In the present work a simple probabilistic (combinatorial) model is formulated to account for a limited ZDC acceptance and to study the impact of this limitation on measured ZDC energy spectra. In this model the numbers $\mathsf n_i$ of {\em detected} events of nucleon multiplicity $i$, are related with the numbers of {\em true} events $\mathsf N_i$. In particular, this model can be applied to forward nucleons emitted in the EMD, where one-nucleon and two-nucleon channels dominate~\cite{Pshenichnov2001}. Due to this dominance it is sufficient to consider only the emission of one, two, three and four nucleons to find the connection between $\mathsf n_1$, $\mathsf n_2$,  $\mathsf n_3$,  $\mathsf n_4$ and $\mathsf N_1$, $\mathsf N_2$,  $\mathsf N_3$, $\mathsf N_4$. These numbers are connected by means of a triangular transformation matrix $\mathsf P$: 
\begin{linenomath*}
\begin{equation}
\left(
\begin{array}{c}
\mathsf n_1\\
\mathsf n_2\\
\mathsf n_3\\
\mathsf n_4\\
\end{array}
\right)=
\left(
\begin{array}{cccc}
\mathsf p_{11} & \mathsf p_{12} &  \mathsf p_{13}  & \mathsf p_{14} \\
        0      & \mathsf p_{22} &  \mathsf p_{23}  & \mathsf p_{24} \\
        0      &         0      &  \mathsf p_{33}  & \mathsf p_{34} \\
        0      &         0      &         0        & \mathsf p_{44} \\                    
\end{array}
\right)
\left(
\begin{array}{c}
\mathsf N_1\\
\mathsf N_2\\
\mathsf N_3\\
\mathsf N_4\\
\end{array}
\right)=\mathsf P
\left(
\begin{array}{c}
\mathsf N_1\\
\mathsf N_2\\
\mathsf N_3\\
\mathsf N_4\\
\end{array}
\right) \ .
\label{trans1234}
\end{equation}
\end{linenomath*}
The diagonal elements of $\mathsf P$ represent the probabilities $\mathsf p_{11}$,...,$\mathsf p_{44}$ to detect exactly the same numbers of forward nucleons as were emitted in events with respective multiplicity $i=1,...,4$. The off-diagonal elements $\mathsf p_{kn}$, $k<n$ represent the probability to detect $k$ nucleons out of $n$ emitted. In ZDC energy spectra low-multiplicity peaks are filled by high-multiplicity events as some of nucleons are lost.  

The most reliable way to obtain $\mathsf p_{kn}$ consists in Monte Carlo modeling of the respective  experimental setup. However, one can assume that the probability $\mathsf p$ to detect a forward nucleon remains the same in low and high multiplicity events. This condition holds when the transverse momentum distribution of forward nucleons has a weak dependence on the event multiplicity. This assumption leads to the binomial distribution of the probabilities with its parameter $\mathsf p$: 
\begin{linenomath*}
\begin{equation}
\mathsf p_{kn}=\binom{n}{k}\mathsf p^{k}(1-\mathsf p)^{n-k} \ . 
\end{equation} 
\end{linenomath*}
\noindent Here the binomial coefficient is defined as $\binom{n}{k}=n!/(n-k)!k!$~. Following this assumption, the transformation matrix is written as:
\begin{linenomath*}
\begin{equation}
\mathsf P=
\left(
\begin{array}{cccc}
\mathsf p & 2\mathsf p(1-\mathsf p) & 3\mathsf p(1-\mathsf p)^2 & 4\mathsf p(1-\mathsf p)^3 \\
        0 & \mathsf p^2 & 3\mathsf p^2(1-\mathsf p) & 6\mathsf p^2(1-\mathsf p)^2 \\
        0      &         0      &  \mathsf p^3 &  4\mathsf p^3(1-\mathsf p) \\
        0      &         0      &         0        &   \mathsf p^4 \\                    
\end{array}
\right)
\label{matrixP}
\end{equation} 
\end{linenomath*}
\noindent Due to a limited ZDC acceptance the detection of multinucleon events is suppressed, while the relative contribution of detected single-nucleon events is enhanced. In order to obtain {\em true} numbers $\mathsf N_i$ of events of each multiplicity, an inverse transformation can be applied to the numbers $\mathsf n_i$ of {\em detected} events: 
\begin{linenomath*}
\begin{equation}
\left(
\begin{array}{c}
\mathsf N_1\\
\mathsf N_2\\
\mathsf N_3\\
\mathsf N_4\\
\end{array}
\right)=
\mathsf P^{-1}
\left(
\begin{array}{c}
\mathsf n_1\\
\mathsf n_2\\
\mathsf n_3\\
\mathsf n_4\\
\end{array}
\right) \ ,
\end{equation}
\end{linenomath*}
\noindent with the following explicit result:
\begin{linenomath*}
\begin{equation}
\left\{
 \begin{array}{ll}
   \mathsf N_1= \cfrac{1}{\mathsf p}\Big(\mathsf n_1-\cfrac{2(1-\mathsf p)}{\mathsf p}\mathsf n_2+\cfrac{3(1-\mathsf p)^2}{\mathsf p^2}\mathsf n_3-\cfrac{4(1-\mathsf p)^3}{\mathsf p^3}\mathsf n_4\Big)\\ 
   \\
   \mathsf N_2= \cfrac{1}{\mathsf p^2}\Big(\mathsf n_2-\cfrac{3(1-\mathsf p)}{\mathsf p}\mathsf n_3+\cfrac{6(1-\mathsf p)^2}{\mathsf p^2}\mathsf n_4  \Big)\\
   \\
   \mathsf N_3= \cfrac{1}{\mathsf p^3}\Big(\mathsf n_3-\cfrac{4(1-\mathsf p)}{\mathsf p}\mathsf n_4  \Big)\\
   \\
   \mathsf N_4= \cfrac{1}{\mathsf p^4}\mathsf n_4\\
                \end{array}
              \right.
              \label{explicit}
\end{equation}
\end{linenomath*}
The set of equations~(\ref{explicit}) can be extended to arbitrary nucleon multiplicity with the maximum number of emitted nucleons denoted as $m$, $i=1,...,m$. This results in the following set of equations: 
\begin{linenomath*}
\begin{equation}
\mathsf N_i= \cfrac{1}{\mathsf p^i}\sum_{j=i}^{m}(-1)^{j-i}\binom{j}{i}\cfrac{(1-\mathsf p)^{j-i}}{\mathsf p^{j-i}}\mathsf n_j \ .
\label{general_form}
\end{equation}
\end{linenomath*}
In other words, the elements of the inverse $\mathsf P^{-1}=\mathsf R$ matrix are calculated as:
\begin{linenomath*}
\begin{equation}
\mathsf r_{kn}=(-1)^{n-k}\binom{n}{k}\frac{(1-\mathsf p)^{n-k}}{\mathsf p^{n}} \ .
\label{inverse_mx_elem}
\end{equation} 
\end{linenomath*}
\noindent Expressions~(\ref{general_form}) or (\ref{inverse_mx_elem}) can be used in estimating the numbers of spectator neutrons and protons emitted in hadronic interactions of nuclei with $m$ approaching the total numbers of neutrons and protons in each of the colliding nuclei.

\section{Acceptance corrections to ZDC energy spectra at the LHC and NICA/MPD}

On the basis of Eqs.~(\ref{trans1234}) and (\ref{matrixP})  one can study the impact of the ZDC acceptance on the energy spectra measured, for example, in the EMD. For this purpose the ZDC spectra for proton and neutron ZDC are modeled below for ultraperipheral $^{208}$Pb--$^{208}$Pb collisions at the LHC at $\sqrt{s_{\rm NN}}$=2.76~TeV ($E_0 = 1380 $~GeV) and $\sqrt{s_{\rm NN}}$=5.02~TeV ($E_0 = 2510$~GeV). The ZDC energy spectra for ultraperipheral $^{197}$Au--$^{197}$Au collisions at $\sqrt{s_{\rm NN}}$=9~GeV ($E_0 = 4.5$~GeV) at future NICA/MPD facility are modeled as well. The rates of emission of given numbers of neutrons and protons are calculated by means of RELDIS model~\cite{Pshenichnov2011,Pshenichnov2001} in all three cases.  As shown in Refs.~\cite{Abelev2012n} and~\cite{Golubeva2005}, the results of this model for 1n-, 2n- and 3n-emission of forward neutrons in the EMD of lead nuclei, respectively, at the LHC and at the CERN SPS agree well with the measured cross sections. A good agreement of RELDIS predictions with the measurements performed at RHIC for $^{197}$Au--$^{197}$Au ultraperipheral collisions has been also reported~\cite{Baltz1996,Chiu2002}. This all gives us confidence in using RELDIS for calculating the ZDC energy spectra from the EMD of lead and gold nuclei in the present work.   

\subsection{ZDC spectra for $^{208}$Pb--$^{208}$Pb collisions at the LHC}

\begin{figure}[!htb]
\begin{centering}
\includegraphics[width=1.0\columnwidth]{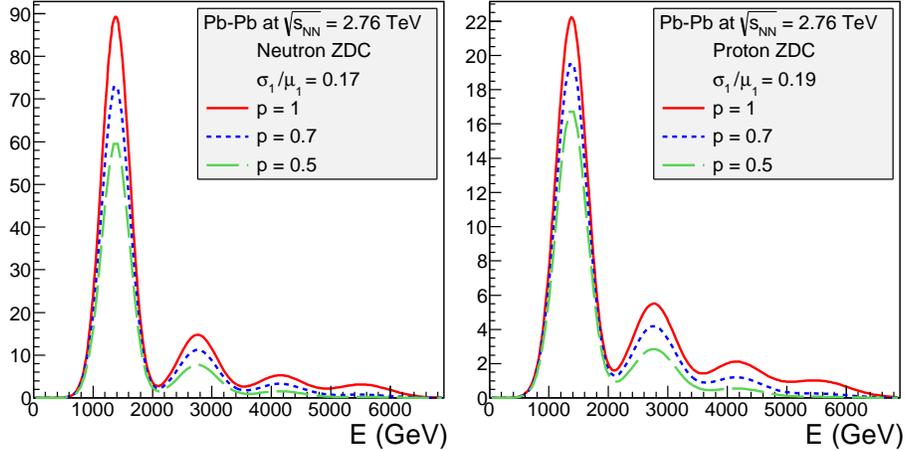}
\caption{Energy distribution (arb. unit) in neutron ZDC (left) and  proton ZDC (right) for ultraperipheral $^{208}$Pb--$^{208}$Pb collisions at the LHC at $\sqrt{s_{\rm NN}}$=2.76~TeV. 
}
\label{fig:LHC_2760}
\end{centering}
\end{figure}

\begin{figure}[!htb]
\begin{centering}
\includegraphics[width=1.0\columnwidth]{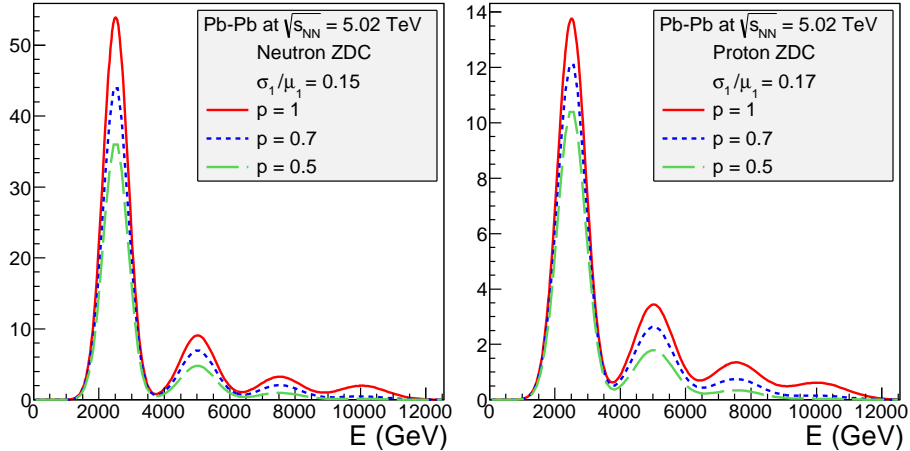}
\caption{Energy distribution (arb. unit) in neutron ZDC (left) and proton ZDC (right) for ultraperipheral $^{208}$Pb--$^{208}$Pb collisions at the LHC at $\sqrt{s_{\rm NN}}$=5.02~TeV.
}
\label{fig:LHC_5020}
\end{centering}
\end{figure}

Following the extrapolations of test beam results of Ref.~\cite{Puddu2007} for the ALICE neutron ZDC, $\sigma_1/\mu_1$  can be estimated as 17\% and 15\% for $E_0=1380$~GeV and $E_0=2510$~GeV, respectively, while for the ALICE proton ZDC it amounts to 19\% and 17\% for these two beam energies. The energy resolution of the neutron ZDC of 20\% directly measured at $E_0=1380$~GeV~\cite{Abelev2012n} is slightly worse. The ALICE experiment is the only experiment at the LHC which is equipped both with the neutron and proton ZDC. The energy spectra in the ALICE ZDC for forward neutrons and protons calculated with the extrapolated parameters for ultraperipheral $^{208}$Pb--$^{208}$Pb collisions are presented in Figs.~\ref{fig:LHC_2760} and~\ref{fig:LHC_5020}.  

The above-mentioned effect of the relative enhancement of single-nucleon peaks with respect to multinucleon ones in the measured spectra is clearly visible in Figs.~\ref{fig:LHC_2760} and \ref{fig:LHC_5020}. Indeed, the heights of 1n and 1p peaks modeled for $\mathsf p=0.5$ are noticeably higher than the half of the heights of the same peaks calculated for the full acceptance of $\mathsf p=1$.  As seen, 1n,...4n, as well as 1p,...4p emission rates in ulraperipheral $^{208}$Pb--$^{208}$Pb collisions at the LHC can be reliably measured by a full acceptance ZDC ($\mathsf p=1$). The impact of a limited ZDC acceptance is basically the same at both collision energies. The options of $\mathsf p=0.5$ and~0.7 make 4n and 4p peaks in the ZDC energy spectra invisible, but 3n and 3p peaks still can be detected. However, 3p peak is essentially melted with $\mathsf p=0.5$. 

\begin{figure}[!htb]
\begin{centering}
\includegraphics[width=1.0\columnwidth]{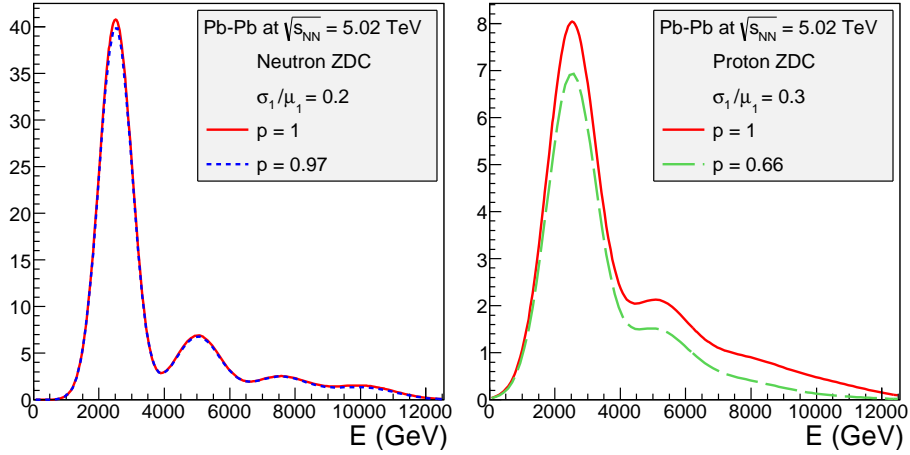}
\caption{Energy distribution (arb. unit) in neutron ZDC (left) and proton ZDC (right) for ultraperipheral $^{208}$Pb--$^{208}$Pb collisions at the LHC at $\sqrt{s_{\rm NN}}$=5.02~TeV in ALICE experiment.
}
\label{fig:LHC_5020_p97_p66}
\end{centering}
\end{figure}

A good ZDC energy resolution is crucial for reliable measurements of the emission of given numbers of forward nucleons. This is illustrated by the ZDC spectra shown in Fig.~\ref{fig:LHC_5020_p97_p66}, which were modeled for a lower resolution in comparison to the spectra shown in Figs.~\ref{fig:LHC_2760} and~\ref{fig:LHC_5020}.  The acceptance and energy resolution were taken as $\mathsf p=0.97$ and $\sigma_1/\mu_1=0.2$ for the neutron ZDC according to Refs.~\cite{Puddu2007,Abelev2012n}. A lower resolution of $\sigma_1/\mu_1=0.3$ was assumed for the proton ZDC and the acceptance of $\mathsf p=0.66$ was taken following Ref.~\cite{Gemme2006}.  

As seen from Fig.~\ref{fig:LHC_5020_p97_p66}, a low resolution of the proton ZDC makes it difficult to identify 2p and 3p channels of the EMD even in the full acceptance measurements. Moreover, 2p peak is noticeably melted with the reduced acceptance of $\mathsf p=0.66$, while the contributions of 3p and 4p EMD channels become indistinguishable from a possible smooth background contribution to the proton ZDC spectra. This demonstrates again the effect of misidentification of multinucleon events as single-nucleon ones in low acceptance measurements.  

\subsection{ZDC spectra for $^{197}$Au--$^{197}$Au collisions at NICA}

\begin{figure}[!htb]
\begin{centering}
\includegraphics[width=1.0\columnwidth]{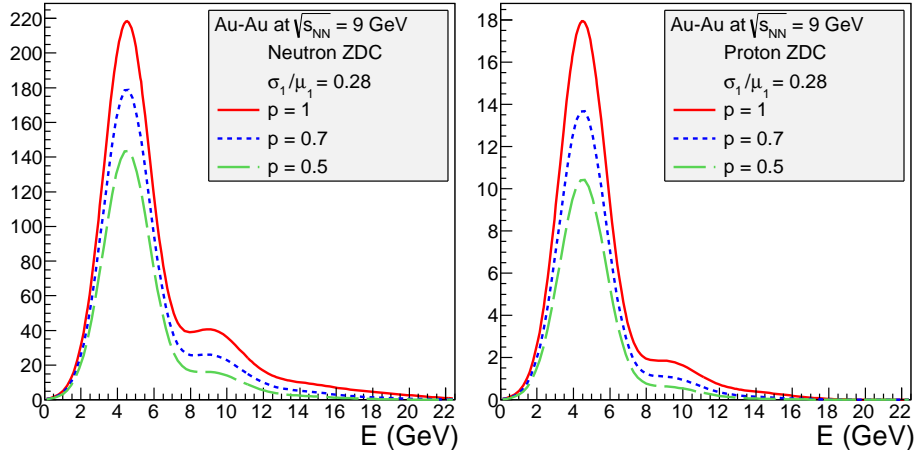}
\caption{Energy distribution (arb. unit) for neutrons (left) and protons (right) in FHCal for $^{197}$Au--$^{197}$Au collisions at NICA/MPD at $\sqrt{s_{\rm NN}}$=9~GeV. 
}
\label{fig:NICA_9}
\end{centering}
\end{figure}

The Forward Hadron Calorimeter (FHCal) designed for MPD experiment at NICA has the energy resolution $\sigma_1/\mu_1$ better than $60\%/\sqrt{E_0}$~\cite{Golubeva2017a}. The width of the first proton peak has been evaluated as $\sigma_1=0.6/\sqrt{E_0}$~GeV, resulting to $\sigma_1/\mu_1=0.28$ at $E_0 = 4.5$~GeV. The energy spectra in FHCal for forward neutrons and protons calculated with these parameters for ultraperipheral $^{197}$Au--$^{197}$Au collisions are presented in Fig.~\ref{fig:NICA_9}. 

NICA provides a lower collision energy compared to the LHC and much less EMD channels are opened at this energy:  
1n (69\%) and 2n (18\%) channels are dominant with only 5\% for 3n.  At NICA 1p, 2p and 3p EMD channels are represented by 5.7\%, 0.9\% and 0.2\% of events, respectively, while protons are not emitted at all in 93\% of EMD events. As seen from Fig.~\ref{fig:NICA_9}, it is difficult to distinguish 3n and 3p peaks in the FHCal energy spectra even in the full acceptance $\mathsf p=1$ measurements. Moreover, 2n and 2p peaks almost disappear with the lowest considered acceptance of $\mathsf p=0.5$. Therefore, it is important to provide a good acceptance for forward nucleons to study the EMD of gold nuclei at NICA or monitor NICA luminosity on the basis of the sum of 1n and 2n EMD channels similarly to the method proposed for the LHC~\cite{Pshenichnov2011}. 

\section{Systematic measurement errors resulting from the uncertainty of the ZDC acceptance}

As a rule, the acceptance of ZDC is estimated as accurate as possible by means of Monte Carlo modelling of the experimental setup. However, the choice of physics models and parameters used in simulations can afect the calculated acceptance and, eventually, the measured rates of emission of forward nucleons. The simple model of the present work makes possible to estimate the uncertainties of measured yields of emission of given numbers of nucleons resulting from the uncertainty of $\mathsf p$. With this purpose $\mathsf N_i$ given by Eqs.~(\ref{general_form}) can be calculated with different values of $\mathsf p$ to characterize the effect of its variation. 

The changes of reconstructed neutron emission rates in the EMD of $^{208}$Pb at $\sqrt{s_{\rm NN}}$=2.76~TeV due to a 10\% reduction of $\mathsf p$ are presented in Table~\ref{tab:n_multiplicity}. The measured fractions of 1n, 2n and 3n emission were taken from Ref.~\cite{Abelev2012n}, but the yield of 4n emission has not been reported in~\cite{Abelev2012n}. In order to supplement the data, the yield of 4n emission was estimated as $\sim 70$\% of the measured 3n emission according to their ratio predicted by RELDIS model~\cite{Pshenichnov2011,Pshenichnov2001}. Since the probability $\mathsf p$ to register a neutron from the EMD in the neutron ZDC has been reported as 0.987~\cite{Abelev2012n}, it is possible to estimate the respective uncorrected yields also listed in Table~\ref{tab:n_multiplicity}. In a hypothetical case of the data reconstruction with a reduced $\mathsf p=0.888$ (--10\%) instead of the true value of 0.987 the resulting yields listed in Table~\ref{tab:n_multiplicity} deviate from the measured yields. The most noticeable difference is found for 2n (+13\%) and 4n (+53\%) yields, while the deviation of 1n yield is rather small (+6\%). As follows from Table~\ref{tab:n_multiplicity}, it is important to measure 4n emission in order to obtain a reliable result for 3n emission. In the case of neglecting 4n emission the 3n yield is reconstructed with a very large error (+42\%). 

\begin{table}[!ht]
\caption{Measured neutron emission fractions (\%) for the single EMD minus mutual EMD of 
$^{208}$Pb at $\sqrt{s_{\rm NN}}$=2.76~TeV\protect\cite{Abelev2012n}, estimated raw data and reconstructed fractions with $\mathsf p=0.987$ and 0.888, with and without 4n contribution. The yield of 4n emission has been estimated as 70\% of the measured 3n emission according to their ratio calculated by RELDIS. The relative differences (in \%) between the data reconstructed with different $\mathsf p$ are given in parentheses. }
\label{tab:n_multiplicity}
\begin{center}
\begin{tabular}{|c|c|c|c|c|}
\hline
\multirow{2}{*}{Channel}  &  Data                     & Raw data         & \multicolumn{2}{|c|}{Reconstructed }  \\
\cline{4-5}
&\protect\cite{Abelev2012n}& $\mathsf p=0.987$ &  $\mathsf p=0.888$ (--10\%)& $\mathsf p=0.888$ (--10\%), w/o 4n \\       
\hline
1nX & $51.5 \pm 0.5 $          & 51.13          &  54.5 (+6\%)   &  54.52  (+6\%)\\
\hline
2nX & $11.6 \pm 0.6 $          & 11.44        &  13.07 (+13\%)  &  12.79 (+10\%)\\
\hline
3nX & $3.6 \pm 0.3$            &  3.59         &  3.41 (--5\%) &  5.12 (+42\%)\\
\hline
4nX & $\sim 2.5$               &  2.37         &  3.82 (+53\%)  &  0. \\
\hline
\end{tabular}
\end{center}
\end{table}

\section{Operation of a ZDC with a low acceptance}

Providing that $i$ forward nucleons are emitted in an event, the probability to have no nucleons detected in this event is $(1-\mathsf p)^i$. Therefore, the probability of detecting at least one of $i$ nucleons is obtained as $1-(1-\mathsf p)^i$. Since events of various multiplicity $i=1,...m$ are possible, the total number of {\em detected} events of any kind is calculated from the numbers of {\em true} events $\mathsf N_i$:
\begin{linenomath*}
\begin{equation}
\sum^{m}_{i=1}\mathsf n_i = \sum^{m}_{i=1} \big(1-(1-\mathsf p)^i \big)\mathsf N_i  \ .   
\end{equation}   
\end{linenomath*}
\noindent A similar equation can be written for the visible cross section $\tilde\sigma$ of detecting at least one nucleon by a ZDC with the acceptance p:
\begin{linenomath*}
\begin{equation}
\tilde\sigma=\mathsf p\sigma(1{\rm NX}) + \mathsf p(2-\mathsf p)\sigma(2{\rm NX})+\mathsf p(3-3\mathsf p+\mathsf p^2)\sigma(3{\rm NX})+ ..... \ .
\label{eq:GeneralCase}
\end{equation}
\end{linenomath*}
\noindent Here $\sigma(1{\rm NX}), \sigma(2{\rm NX}), \sigma(3{\rm NX})$ etc. denote the cross sections of emission of one, two, three etc., forward nucleons. 

One can consider a special case of a ZDC with a low acceptance. In this case many forward nucleons remain undetected.
In the limit of $\mathsf p\rightarrow 0$ Eq.~(\ref{eq:GeneralCase}) is reduced to:
\begin{linenomath*}
\begin{equation}
\tilde\sigma=\mathsf p\left(\sigma(1{\rm NX}) + 2\sigma(2{\rm NX})+ 3\sigma(3{\rm NX})\right)+ ... =\mathsf p\sigma_{inc}\ ,
\label{eg:inclusve}
\end{equation}
\end{linenomath*}
\noindent where $\sigma_{inc}$ is the inclusive cross section of nucleon emission. This means that  $\sigma_{inc}$ can be estimated from the visible cross section as $\sigma_{inc}=\tilde\sigma/\mathsf p$ when $\mathsf p$ is very low. Alternatively, providing that $\sigma_{inc}$ is known from other measurements or predicted by a theory and also $\mathsf p$ is obtained from Monte Carlo modeling, the collider luminosity monitoring can be performed on the basis of the visible cross section $\tilde\sigma$. In summary, it is difficult to measure the yields of partial channels with given numbers of forward nucleons by means of the ZDC with a limited acceptance. However, such a device still can be used for the measurements of the inclusive nucleon emission cross section. 

\section{Conclusions}
 
Monte Carlo modelling of the ZDC used by ALICE, NICA or other heavy-ion collision experiments is out of the scope of the present work. The results of the simple probabilistic model formulated here can be less accurate than the Monte Carlo results, but still demonstrate general features of detecting multinucleon EMD events by ALICE and NICA. The present study shows that 1n,...4n, as well as 1p,...4p emission rates in ulraperipheral $^{208}$Pb--$^{208}$Pb collisions at the LHC can be reliably measured by a full acceptance ZDC ($\mathsf p=1$) providing that the ZDC energy resolution is better than 20\%. However, with $\mathsf p=0.7$ and the same resolution 3n and 3p peaks still remain visible in the ZDC energy spectra, but 4n and 4p peaks disappear. As found, the performance of the ZDC  essentially deteriorates when both the resolution and acceptance are reduced. In particular, 2p peak in the proton ZDC spectrum is noticeably melted with the reduced acceptance of $\mathsf p=0.66$, while the contributions of 3p and 4p EMD channels become indistinguishable from a possible smooth background contribution to the proton ZDC spectra. In all cases of a limited ZDC acceptance ($\mathsf p<1$) it is important to use Eqs.~(\ref{general_form}) derived in the present work to extract the rates of emission of given numbers of forward nucleons from the respective measured rates. 

A very small acceptance of $\mathsf p\rightarrow 0$ makes unreliable the measurements of individual channels of emission of forward nucleons. However, a low acceptance ZDC still can be used for the measurements of the inclusive nucleon emission cross section and for the purpose of the collider luminosity monitoring.   

The modeled energy spectra of forward neutrons and protons from the EMD at future NICA/MPD facility also demonstrate the suppression of multinucleon channels due to a limited ZDC acceptance. Therefore, it is crucial to build a full-acceptance ZDC to study the EMD of gold nuclei at NICA or monitor NICA luminosity on the basis of the sum of 1n and 2n EMD channels similarly to the method proposed for the LHC.

\section*{References}

\bibliography{ZDCPerformance}

\begin{thebibliography}{10}
\expandafter\ifx\csname url\endcsname\relax
  \def\url#1{\texttt{#1}}\fi
\expandafter\ifx\csname urlprefix\endcsname\relax\def\urlprefix{URL }\fi
\expandafter\ifx\csname href\endcsname\relax
  \def\href#1#2{#2} \def\path#1{#1}\fi

\bibitem{Golubeva2013}
M.~B. Golubeva, F.~F. Guber, A.~P. Ivashkin, A.~Y. Isupov, A.~B. Kurepin, A.~G.
  Litvinenko, E.~I. Litvinenko, I.~I. Migulina, V.~F. Peresedov,
  {Nuclear-nuclear collision centrality determination by the spectators
  calorimeter for the MPD setup at the NICA facility}, Phys. At. Nucl. 76~(1)
  (2013) 1--15.
\newblock \href {http://dx.doi.org/10.1134/S1063778812120046}
  {\path{doi:10.1134/S1063778812120046}}.

\bibitem{Golubeva2017}
M.~B. Golubeva, A.~P. Ivashkin, A.~B. Kurepin, {Study of nuclear fragmentation
  at MPD/NICA}, EPJ Web Conf. 138 (2017) 11001.
\newblock \href {http://dx.doi.org/10.1051/epjconf/201713811001}
  {\path{doi:10.1051/epjconf/201713811001}}.

\bibitem{Adler2001}
C.~Adler, A.~Denisov, E.~Garcia, M.~Murray, H.~Stroebele, S.~White, {The RHIC
  zero degree calorimeters}, Nucl. Instruments Methods Phys. Res. Sect. A
  Accel. Spectrometers, Detect. Assoc. Equip. 470~(3) (2001) 488--499.
\newblock \href {http://dx.doi.org/10.1016/S0168-9002(01)00627-1}
  {\path{doi:10.1016/S0168-9002(01)00627-1}}.

\bibitem{Arnaldi2006}
R.~Arnaldi, E.~Chiavassa, C.~Cicalo, P.~Cortese, A.~{De Falco}, G.~Dellacasa,
  N.~{De Marco}, A.~Ferretti, M.~Gallio, R.~Gemme, A.~Masoni, P.~Mereu,
  A.~Musso, C.~Oppedisano, A.~Piccotti, F.~Poggio, G.~Puddu, E.~Scomparin,
  S.~Serci, E.~Siddi, G.~Travaglia, G.~Usai, E.~Vercellin, {The Neutron Zero
  Degree Calorimeter for the ALICE experiment}, Nucl. Instruments Methods Phys.
  Res. Sect. A Accel. Spectrometers, Detect. Assoc. Equip. 564~(1) (2006)
  235--242.
\newblock \href {http://dx.doi.org/10.1016/j.nima.2006.03.044}
  {\path{doi:10.1016/j.nima.2006.03.044}}.

\bibitem{Puddu2007}
G.~Puddu, R.~Arnaldi, E.~Chiavassa, C.~Cical{\'{o}}, P.~Cortese, A.~{De Falco},
  G.~Dellacasa, A.~Ferretti, M.~Floris, M.~Gagliardi, M.~Gallio, R.~Gemme,
  G.~Locci, A.~Masoni, P.~Mereu, A.~Musso, C.~Oppedisano, A.~Piccotti,
  F.~Poggio, E.~Scomparin, S.~Serci, E.~Siddi, D.~Stocco, G.~Usai,
  E.~Vercellin, F.~Yermia, {The zero degree calorimeters for the ALICE
  experiment}, Nucl. Instruments Methods Phys. Res. Sect. A Accel.
  Spectrometers, Detect. Assoc. Equip. 581~(1-2) (2007) 397--401.
\newblock \href {http://dx.doi.org/10.1016/j.nima.2007.08.013}
  {\path{doi:10.1016/j.nima.2007.08.013}}.

\bibitem{Gemme2009}
R.~Gemme, R.~Arnaldi, E.~Chiavassa, C.~Cicalo, P.~Cortese, A.~{De Falco},
  G.~Dellacasa, N.~{De Marco}, A.~Ferretti, M.~Floris, M.~Gagliardi, M.~Gallio,
  G.~Luparello, A.~Masoni, P.~Mereu, A.~Musso, C.~Oppedisano, A.~Piccotti,
  F.~Poggio, G.~Puddu, E.~Scomparin, S.~Serci, E.~Siddi, D.~Stocco, G.~Usai,
  E.~Vercellin, {Commissioning and calibration of the Zero Degree Calorimeters
  for the ALICE experiment}, Nucl. Phys. B - Proc. Suppl. 197~(1) (2009)
  211--214.
\newblock \href {http://dx.doi.org/10.1016/j.nuclphysbps.2009.10.069}
  {\path{doi:10.1016/j.nuclphysbps.2009.10.069}}.

\bibitem{White2010}
S.~White, {The ATLAS zero degree calorimeter}, Nucl. Instruments Methods Phys.
  Res. Sect. A Accel. Spectrometers, Detect. Assoc. Equip. 617~(1-3) (2010)
  126--128.
\newblock \href {http://dx.doi.org/10.1016/j.nima.2009.09.120}
  {\path{doi:10.1016/j.nima.2009.09.120}}.

\bibitem{Grachov2011}
O.~Grachov, M.~Murray, J.~Wood, Y.~Onel, S.~Sen, T.~Yetkin, {Commissioning of
  CMS zero degree calorimeter using LHC beam}, J. Phys. Conf. Ser. 293 (2011)
  012040.
\newblock \href {http://dx.doi.org/10.1088/1742-6596/293/1/012040}
  {\path{doi:10.1088/1742-6596/293/1/012040}}.

\bibitem{Abelev2013o}
B.~Abelev, et~al., {Centrality determination of Pb-Pb collisions at
  $\sqrt{s_{NN}}=2.76$ TeV with ALICE}, Phys. Rev. C 88~(4) (2013) 044909.
\newblock \href {http://dx.doi.org/10.1103/PhysRevC.88.044909}
  {\path{doi:10.1103/PhysRevC.88.044909}}.

\bibitem{Baltz1996}
A.~J. Baltz, M.~J. Rhoades-Brown, J.~Weneser, {Heavy-ion partial beam lifetimes
  due to Coulomb induced processes}, Phys. Rev. E 54~(4) (1996) 4233--4239.
\newblock \href {http://dx.doi.org/10.1103/PhysRevE.54.4233}
  {\path{doi:10.1103/PhysRevE.54.4233}}.

\bibitem{Pshenichnov2011}
I.~Pshenichnov, {Electromagnetic excitation and fragmentation of
  ultrarelativistic nuclei}, Phys. Part. Nucl. 42~(2) (2011) 215--250.
\newblock \href {http://dx.doi.org/10.1134/S1063779611020067}
  {\path{doi:10.1134/S1063779611020067}}.

\bibitem{Pshenichnov2001}
I.~A. Pshenichnov, J.~P. Bondorf, I.~N. Mishustin, A.~Ventura, S.~Masetti,
  {Mutual heavy ion dissociation in peripheral collisions at ultrarelativistic
  energies}, Phys. Rev. C 64~(2) (2001) 249031--2490319.
\newblock \href {http://arxiv.org/abs/0101035} {\path{arXiv:0101035}}, \href
  {http://dx.doi.org/10.1103/PhysRevC.64.024903}
  {\path{doi:10.1103/PhysRevC.64.024903}}.

\bibitem{Abelev2012n}
B.~Abelev, et~al., {Measurement of the Cross Section for Electromagnetic
  Dissociation with Neutron Emission in Pb-Pb Collisions at
  $\sqrt{s_{NN}}=2.76$ TeV}, Phys. Rev. Lett. 109~(25) (2012) 252302.
\newblock \href {http://arxiv.org/abs/1203.2436} {\path{arXiv:1203.2436}},
  \href {http://dx.doi.org/10.1103/PhysRevLett.109.252302}
  {\path{doi:10.1103/PhysRevLett.109.252302}}.

\bibitem{Gemme2006}
R.~Gemme, \href{http://cds.cern.ch/record/1322418}{{Study of the ALICE ZDC
  detector performance}}, Ph.D. thesis, Turin University (2006).
\newline\urlprefix\url{http://cds.cern.ch/record/1322418}

\bibitem{Awes1989}
T.~Awes, C.~Baktash, R.~Cumby, R.~Ferguson, A.~Franz, T.~Gabriel,
  H.~Gustafsson, H.~Gutbrod, J.~Johnson, B.~Kolb, I.~Lee, F.~Obenshain,
  A.~Oskarsson, I.~Otterlund, S.~Persson, F.~Plasil, A.~Poskanzer, H.~Ritter,
  H.~Schmidt, S.~Sorensen, G.~Young, {The mid-rapidity calorimeter for the
  relativistic heavy-ion experiment WA80 at CERN}, Nucl. Instruments Methods
  Phys. Res. Sect. A Accel. Spectrometers, Detect. Assoc. Equip. 279~(3) (1989)
  479--502.
\newblock \href {http://dx.doi.org/10.1016/0168-9002(89)91295-3}
  {\path{doi:10.1016/0168-9002(89)91295-3}}.

\bibitem{Karpechev2014}
E.~V. Karpechev, I.~Pshenichnov, T.~L. Karavicheva, A.~Kurepin, M.~B. Golubeva,
  F.~F. Guber, A.~I. Maevskaya, A.~I. Reshetin, V.~V. Tiflov, N.~S. Topilskaya,
  P.~Cortese, G.~Dellacasa, R.~Arnaldi, N.~{De Marco}, A.~Ferretti, M.~Gallio,
  A.~Musso, C.~Oppedisano, A.~Piccotti, E.~Scomparin, E.~Vercellin, C.~Cicalo,
  G.~Puddu, E.~Siddi, P.~Szymanski, I.~Efthymiopoulos, {Emission of forward
  neutrons by 158A GeV indium nuclei in collisions with Al, Cu, Sn and Pb},
  Nucl. Phys. A 921 (2014) 60--84.
\newblock \href {http://dx.doi.org/10.1016/j.nuclphysa.2013.11.003}
  {\path{doi:10.1016/j.nuclphysa.2013.11.003}}.

\bibitem{Golubeva2005}
M.~B. Golubeva, F.~F. Guber, T.~L. Karavicheva, E.~V. Karpechev, A.~B. Kurepin,
  A.~I. Maevskaya, I.~A. Pshenichnov, A.~I. Reshetin, K.~A. Shileev, V.~V.
  Tiflov, N.~S. Topilskaya, P.~Szymanski, I.~Efthymiopoulos, L.~Gatignon,
  P.~Cortese, G.~Dellacasa, R.~Arnaldi, N.~{De Marco}, A.~Ferretti, M.~Gallio,
  A.~Musso, C.~Oppedisano, A.~Piccotti, E.~Scomparin, E.~Vercellin, C.~Cicalo,
  G.~Puddu, E.~Siddi, {Neutron emission in electromagnetic dissociation of
  ultrarelativistic Pb ions}, Phys. Rev. C - Nucl. Phys. 71~(2) (2005) 1--10.
\newblock \href {http://dx.doi.org/10.1103/PhysRevC.71.024905}
  {\path{doi:10.1103/PhysRevC.71.024905}}.

\bibitem{Chiu2002}
M.~Chiu, A.~Denisov, E.~Garcia, J.~Katzy, A.~Makeev, M.~Murray, S.~White,
  {Measurement of Mutual Coulomb Dissociation in $\sqrt{s_{NN}}=130$~GeV Au+Au
  Collisions}, Phys. Rev. Lett. 89~(1) (2002) 1--5.
\newblock \href {http://arxiv.org/abs/0109018} {\path{arXiv:0109018}}, \href
  {http://dx.doi.org/10.1103/PhysRevLett.89.012302}
  {\path{doi:10.1103/PhysRevLett.89.012302}}.

\bibitem{Golubeva2017a}
M.~Golubeva, F.~Guber, A.~Ivashkin, A.~Izvestnyy, M.~Kapishin, A.~Kurepin,
  A.~Litvinenko, E.~Litvinenko, I.~Migulina, S.~Morozov, P.~Parfenov,
  V.~Peresedov, O.~Petukhov, I.~Selyuzhenkov, I.~Svintsov, A.~Taranenko,
  A.~Zinchenko,
  \href{http://mpd.jinr.ru/wp-content/uploads/2017/08/MPD_TDR_FHCal_v9_1.pdf}{{Forward
  Hadron Calorimeter (FHCal). Technical Design Report for the MPD Experiment}},
  Tech. rep., JINR, Dubna (2017).
\newline\urlprefix\url{http://mpd.jinr.ru/wp-content/uploads/2017/08/MPD_TDR_FHCal_v9_1.pdf}

\end{thebibliography}

\end{document}